\documentclass[3p,floatfix,showpacs,tightenlines,showkeys,superscriptaddress,amsmath,amssymb,nofootinbib]{elsarticle}

\usepackage{amssymb,amsbsy,epsfig,color,graphicx}
\usepackage{color}
\usepackage{subfigure}
\usepackage{[longtable}
\usepackage{array}
\usepackage{dcolumn}   
\usepackage{cellspace}
\usepackage{mathtools}
\usepackage{amstext}
\usepackage{amssymb}
\usepackage{stmaryrd}
\usepackage{stackrel}
\usepackage{graphicx}
\usepackage{esint}
\usepackage[utf8]{inputenc}
\usepackage{blindtext}
\usepackage{float}
\restylefloat{table}
\usepackage{booktabs}

\usepackage{etoolbox} 
\usepackage{lipsum} 
\usepackage[capitalize]{cleveref}

\journal{Nuclear Physics B}

\makeatletter

\appto{\appendix}{%
	\@ifstar{\def\theequation@prefix{A.}}%
	{}%
}
\makeatother

\begin{document}

\begin{frontmatter}

\title{Quantum solitonic wave-packet  of a meso-scopic system in singularity free gravity}

\author[first,second,third]{Luca Buoninfante}
\ead{lbuoninfante@sa.infn.it}

\author[first,second]{Gaetano Lambiase}
\ead{lambiase@sa.infn.it}

\author[third,fourth]{Anupam Mazumdar}

\ead{anupam.mazumdar@rug.nl}

\address[first]{Dipartimento di Fisica "E.R. Caianiello", Universit\`a di Salerno, I-84084 Fisciano (SA), Italy}
\address[second]{INFN - Sezione di Napoli, Gruppo collegato di Salerno, I-84084 Fisciano (SA), Italy}
\address[third]{Van Swinderen Institute, University of Groningen, 9747 AG, Groningen, The Netherlands}
\address[fourth]{Kapteyn Astronomical Institute, University of Groningen, 9700 AV Groningen, The Netherlands}

\begin{abstract}
In this paper we will discuss how to localise a quantum wave-packet due to self-gravitating meso-scopic object by taking into account gravitational self-interaction in the Schr\"odinger equation beyond General Relativity. In particular, we will study soliton-like solutions in infinite derivative ghost free theories of
gravity, which resolves the gravitational $1/r$ singularity in the potential. We will show a unique feature that the quantum spread of such a gravitational system is larger than that of the Newtonian gravity, therefore enabling us a window of opportunity to test classical and quantum properties of such theories of gravity in the near future at a table-top experiment.
\end{abstract}


\end{frontmatter}


\section{Introduction and motivations}

Einstein's general relativity (GR) has been widely accepted theory of gravity on large scales and late times, i.e. in the infrared (IR) regime, where it has been tested to a very high precision~\cite{-C.-M.}. The recent detection of gravitational waves from binary Blackholes has matched its predictions extremely well with numerical simulations~\cite{-B.-P.}. Despite the great success of GR, there are problems at short distances and small time scales, the theory allows Blackhole and Cosmological singularities in the ultraviolet (UV) regime. In addition, there is an open question - to what extent Einstein's GR is valid in the UV? In fact, the inverse-square law of Newton's potential has been tested only up to $5.6\times10^{-5}$ meters \cite{-D.-J.}. This means that any modification from the Newtonian potential can occur in the vast desert of scales spanning about more than 30 orders of magnitude, i.e from $0.004$ eV to the Planck scale $M_{p}\sim10^{19}\,\text{GeV}$ in $4$ spacetime dimensions. 

This provides us an ample motivation to explore gravitation beyond GR, and put such theories on test in a laboratory, even at table-top experiment. A very well-known interesting observation is that any macroscopic object naturally provide a scale for localising its own quantum wave packet.
For instance, in the Newtonian gravitational potential, a quantum spread of a wave packet has been computed by Di\'osi, which is {\it solely} given by the Newton constant, $G= 1/M_p^2$, and the mass, $m$, of a macroscopic object by looking for ground-state solutions of minimal energy. An approximate solitonic solution has been found due to self-gravitating potential which tries to confine the wave packet, while quantum effects will try to de-localize the wave packet. An optimum quantum spread, $\sigma_{\scriptscriptstyle N}$, of the wave packet can be found by minimising the energy, and it is  given by~\cite{Diosi:1988uy}:
\begin{equation}
\sigma_{{\scriptscriptstyle N}}=\frac{3}{2}\sqrt{\frac{\pi}{2}}\left(\frac{1}{Gm^{3}}\right)\approx\frac{1}{Gm^{3}}.\label{eq:13}
\end{equation}
Indeed, this quantum spread is very tiny for a massive object, i.e. of the order of $\sigma_{{\scriptscriptstyle N}}\sim {\cal O}(10^{-5})$ m  for 
$m\sim {\cal O}(10^{-18})$ kg.  Therefore, measuring such a spread of a wave packet provides a simple way to constrain theories of gravity beyond GR in a table-top experiment.

The aim of this paper will be to show explicitly that in theories beyond GR, it is possible to constrain the new scale of gravitation by quantifying the solitonic wave packet for a massive object and compare the results with that of the Newtonian case. In particular, we will be studying a particular modification of GR, where the gravitational force vanishes in the UV limit, i.e. $F_g \rightarrow 0$ as $ r\rightarrow 0$, where $F_g$ stands for the force between particles separated by the distance $r$. Such a theory has recently been advocated to ameliorate gravity in the UV~\cite{Biswas:2011ar}.

The result in Eq. \eqref{eq:13} was obtained in Ref. \cite{Diosi:1988uy} by working in the semi-classical approach where gravity is treated as a classical interaction, and the matter component is quantized. However, in this manuscript we will make a more general treatment analyzing both cases of classical and quantized gravity, and discuss the experimental testability of the models in both directions. We will see that in the first case one is able to study the dynamics of self-gravitating one-particle state, as for example elementary particles or mass center of single molecules, by describing the coupling between gravity and quantum matter through a semi-classical approach. While in the second case, one can study the dynamics of a condensate in a mean-field approximation by taking into account all internal mutual gravitational interaction which contribute to the self-potential.

It has been known for a while that a quadratic curvature gravity ($4$ derivatives in the metric) is a renormalizable theory of gravity, but contains massive spin-2 {\it ghost}, signaling instability in the vacuum and therefore lacking predictions~\cite{-K.-S.}. In order to resolve this {\it ghost} problem, we require {\it infinite derivative theories of gravity} (IDG) as pointed out in Ref.~\cite{Biswas:2011ar}. The most general quadratic curvature {\it covariant} action with {\it infinite derivatives} has been constructed in $4$ dimensions, around constant curvature backgrounds, see~\cite{Biswas:2011ar,Biswas:2016etb}: 
\begin{equation}
S=  \displaystyle\frac{1}{16\pi G} \displaystyle\int d^{4}x\sqrt{-g}\left[{ \mathcal{R}}+\alpha\left(\mathcal{R}\mathcal{F}_{1}(\boxempty_{s})\mathcal{R}\right.\right.
\left.\left.+\mathcal{R}_{\mu\nu}\mathcal{F}_{2}(\boxempty_{s})\mathcal{R}^{\mu\nu}+\mathcal{R}_{\mu\nu\rho\sigma}\mathcal{F}_{3}(\boxempty_{s})\mathcal{R}^{\mu\nu\rho\sigma}\right)\right], \label{eq:1}
\end{equation}
where $\alpha $ is a dimensionful coupling,   $\boxempty_{s}\equiv\boxempty/M_{s}^{2}$, where $M_s$ is the new scale of gravitation, i.e. $0.004{\rm eV} \leq M_s\leq 10^{19}$~GeV. The d'Alembertian operator is given by: $\boxempty\equiv g^{\mu\nu}\nabla_{\mu}\nabla_{\nu}$, where $\mu,~\nu=0,1,2,3$, and we take the convention $(-,+,+,+)$ for the metric signature. The three form factors are very similar to {\it pion} form factors in strong interaction, and they are reminiscence to derivative nature of interaction depicted by the massless nature of gravity, i.e. they contain infinite order covariant derivatives and are analytic functions of $\boxempty$, $\mathcal{F}_{i}(\boxempty_{s})=\stackrel[n=0]{\infty}{\sum}f_{i,n}\left(\boxempty_{s}\right)^{n}$. These form factors have already been constrained by the general covariance, in the IR the form factors should be such that they match the predictions of GR, and through-out from IR to UV, they should not introduce any new dynamical degrees of freedom, i.e. the graviton remains massless and transverse-traceless, therefore the coefficients $f_{i,n}$ are all fixed.
Around Minkowski background, they follow a simple relationship given by $2{\cal F}_1(\boxempty_s)+ {\cal F}_2(\boxempty_s) + 2{\cal F}_3(\boxempty_s)=0$~\cite{Biswas:2011ar}. In fact, around a constant curvature background, we can treat ${\cal F}_3=0$ up to quadratic order in the metric perturbation, without loss of generality. It is sufficient to study the quadratic in curvature gravitational action, since we are studying the tree-level gravitational potential.

The new scale of physics, $M_s$, also signifies the scale of non-locality~\cite{-Yu.-V.,Tomboulis,Tseytlin,Siegel,Modesto,Biswas:2005qr}, where the gravitational interaction in this class of theory becomes non-local, see~\cite{Talaganis:2014ida}.  Furthermore, it has been argued that 
the above action becomes UV finite beyond 1-loop~\cite{-Yu.-V.,Tomboulis,Modesto,Talaganis:2014ida}. In this regard, we will be exploring for the {\it first time} quantum localization of a wave-packet in such non-local theories of gravity. Classically, such theories can resolve cosmological singularity as pointed out in~\cite{Biswas:2005qr,Biswas:2011ar,Koshelev:2012qn}, and possibly even blackhole singularity~\cite{Koshelev:2017bxd}, due to the fact that the physical effect of non-locality can be spread out on macroscopic scale in spacetime.

In the following section we will find the fundamental equations describing the dynamics of a quantum meso-scopic system whose self-gravitational interaction is taken into account. 

\section{Dyamics of a self-gravitating system in infinite derivative gravity} 

We now wish to study the dynamics of a self-gravitating quantum system in a low-energy regime, with gravitational interaction described by IDG. We will distinguish 1) the case in which the coupling to quantum matter is given through the expectation value of the stress-energy tensor operator, so that gravity can be treated as a classical interaction, and 2) and the case where gravity is directly coupled to the quantum matter stress-energy tensor operator, such that also the gravitational field has to be quantized. We will obtain the same non-linear Schr\"odinger equation which has two completely different interpretations in the two cases: in 1) it is seen as a fundamental equation describing the dynamics of a self-gravitating one-particle system; while in 2) it is derived as the large $N$ limit ($N\rightarrow\infty$) of a linear Schr\"odinger equation for an $N$-particle state by applying a mean-field approximation, so that the self-gravitational potential will be given by the sum of an infinite number of mutual gravitational interaction-contributions.  

\subsection{Semi-classical approach}

We start considering a semi-classical approach to IDG described by the action in Eq. (\ref{eq:1}), and work in the non-relativistic and in the weak-field regime. The starting point is the field equations with a semi-classical source term:
\begin{equation}
P_{\mu\nu}\equiv G_{\mu\nu}+G^{(q)}_{\mu\nu}=8\pi G\left\langle \psi\left|\hat{\tau}_{\mu\nu}\right|\psi\right\rangle ,\label{eq:3}
\end{equation}
where $G_{\mu\nu}$ is the usual Einstein tensor and $G^{(q)}_{\mu\nu}$ is the contribution to the field equation arising from the quadratic curvature terms in the IDG action; see Ref. \cite{Biswas:2013cha} for the explicit expression and details. The right hand side of the above expression contains the expectation value of the quantised energy-momentum operator $\hat \tau_{\mu\nu}$ in the quantum state 
$\left|\psi\right>$. In the semi-classical approach matter is quantized in a fully classical background, as pointed out in Refs.~\cite{M=0000F8ller,Rosenfeld}. The debate on the validity of semi-classical approach is still an open issue, see~\cite{eppley}, which would warrant further experimental evidences~\cite{mattingly}. In particular, 
we are interested in the linear regime of Eq. (\ref{eq:3}), which can be obtained by expanding the spacetime metric around the Minkowski background, $g_{\mu\nu}=\eta_{\mu\nu}+h_{\mu\nu}$, and neglecting higher order terms in the perturbation $\mathcal{O}(h_{\mu\nu}^{2})$. Moreover, by  imposing the DeDonder gauge, $\partial^{\mu}\left(h_{\mu\nu}-1/2\eta_{\mu\nu}h\right)=0,$ we can now show that the semi-classical linearized equations are given by~\cite{Biswas:2011ar,Biswas:2013cha}:
\begin{equation}
 a(\boxempty_{s})\boxempty h_{\mu\nu}={ -16\pi G\left(\left\langle \psi\left|\hat{\tau}_{\mu\nu}\right|\psi\right\rangle -\frac{1}{2}\eta_{\mu\nu}\left\langle \psi\left|\hat{\tau}\right|\psi\right\rangle \right)}
\label{eq:4}
\end{equation}
where $h\equiv\eta^{\rho\sigma}h_{\rho\sigma}$ and $\hat{\tau}\equiv\eta^{\rho\sigma}\hat{\tau}_{\rho\sigma}$.
Note that the perturbation $h_{\mu\nu}$ is a classical field, i.e.
	it is not quantized in the semi-classical approach.
The coefficient $a(\boxempty_{s})$ is defined in terms of \cite{Biswas:2011ar}:
\begin{equation}
a(\boxempty_{s})=  { 1- \frac{1}{2} \alpha\mathcal{F}_{2}(\boxempty_{s})\boxempty}\,, 
\label{eq:5}
\end{equation}
and $2{\cal F}_1(\boxempty_s)+{\cal F}_2(\boxempty_s)=0$.
Note, that in the linear approximation we are working with, $\hat{\tau}_{\mu\nu}$ acts as the quantized energy-momentum tensor operator in a flat spacetime. 
The coefficient $a(\boxempty_{s})$ is not an arbitrary function as we had discussed above.  By demanding that the gravity remains massless and does not introduce any new dynamical degrees of freedom, the function $a(\boxempty_{s})$ should be {\it exponential of an entire function}~\cite{Tomboulis,Biswas:2011ar}. Such functions do not introduce any new poles in the propagator, and therefore no new dynamical degrees of freedom other than the massless graviton. One simple choice is 
\footnote{One way to show that the choice $a(-k^{2})=e^{k^{2}/M_{s}^{2}}$
	(in the momentum space) does not introduce any additional gravitational
	degrees of freedom is to consider the poles in the propagator. As
	shown in Ref. \cite{Biswas:2011ar,Biswas:2013kla,Buoninfante} the propagator corresponding
	to the action around Minkowski spacetime  is given by $\Pi(-k^2)=(1/a(-k^2))\left[\mathcal{P}^{2}/k^{2}-\mathcal{P}^{0}/2k^{2}\right]$,
	where $\mathcal{P}^{2}$ and $\mathcal{P}^{0}$ are the so called
	spin projection operators along the spin-$2$ and spin-$0$ components,
	respectively. For the choice $a(-k^{2})=e^{k^{2}/M_{s}^{2}}$, there are no additional poles in the propagator, 
	where the GR propagator is given by: $\Pi_{{\scriptscriptstyle GR}}=\mathcal{P}^{2}/k^{2}-\mathcal{P}_{s}^{0}/2k^{2}$.}$^{,}$\footnote{The exponential choice $e^{-\Box /{M_{s}^{2}}}$ is made to ensure UV suppression in the propagator. Indeed, the corresponding propagator in such non-local field theory turns out to be exponentially suppressed for both time-like and space-like momentum-exchange, as it can be shown once it is dressed by summing up all one-particle irreducible diagrams. For example, see Refs. \cite{okada,talag-scatt}.}:
\begin{equation}
a(\boxempty_{s})=e^{-\boxempty/M_s^{2}}.\label{eq:6}
\end{equation}
In fact other choices of {\it entire function} can be made without any loss of generality, as they provide similar  UV and IR behaviour as
pointed out in Ref.~\cite{Edholm:2016hbt}, i.e. the gravitational potential, $\Phi \rightarrow {\rm const.}$ for $r < 2/M_s$ in the UV, therefore the force vanishes in this regime, while in the IR for $r\geq 2/M_s$, the gravitational potential yields $\Phi \sim 1/r$, as expected in the case of GR and in Newtonian gravity~\cite{Biswas:2011ar,Koshelev:2017bxd}.

We now wish to solve Eq. (\ref{eq:4}) in the weak-gravitational field and static spacetime limit, $\hat{\tau}\simeq-\hat{\tau}_{00},$ $\boxempty\simeq\nabla^{2}$ and $\partial_{0}h_{\mu\nu}\simeq0$, for the metric:
\begin{equation}
ds^{2}=-\left(1+2\Phi\right)dt^{2}+\left(1-2\Phi\right)d\vec{x}^{2}\,,
\end{equation}
where $d\vec x^2= dx^2+dy^2+dz^2$, and the gravitational potential, $\left|\Phi\right|\leq 1$, satisfies the differential equation:
\begin{equation}
a\left(\nabla_{s}^{2}\right)\nabla^{2}\Phi=4\pi G\left\langle \psi\left|\hat{\tau}_{00}\right|\psi\right\rangle ,\label{eq:7}
\end{equation}
whose solution is given by:
\begin{equation}
\Phi[\psi](\vec{x})=-G\int d^{3}x'\frac{{\rm Erf}\left(\frac{M_{s}}{2}\left|\vec{x}'-\vec{x}\right|\right)}{\left|\vec{x}'-\vec{x}\right|}\left\langle \psi\left|\hat{\rho}(\vec{x}')\right|\psi\right\rangle ,\label{eq:8}
\end{equation}
where we have used $\hat{\tau}_{00}=\hat{\rho}$.  

We can now compute the semi-classical interaction Hamiltonian, $\!\hat{H}_{\text{int}}=-\frac{1}{2}\int \!d^{3}xh_{\mu\nu}\hat{\tau}^{\mu\nu}$:
\begin{equation}
\begin{array}{rr}
\!\hat{H}_{\text{int}}=\displaystyle -G\int \!d^{3}xd^{3}x'\frac{{\rm Erf}\left(\frac{M_{s}}{2}\left|\vec{x}'-\vec{x}\right|\right)}{\left|\vec{x}'-\vec{x}\right|}\left\langle \psi\left|\hat{\rho}(\vec{x}')\right|\psi\right\rangle\hat{\rho}(\vec{x}),
\end{array}
\label{eq:9}
\end{equation}
where we have used the fact that  $\hat \tau^{00}$ provides the dominant contribution, and $h_{00}=-2\Phi$.
We can now obtain a one-particle Schr\"odinger equation, with a non-linear term that takes into account the gravitational self-interaction~\footnote{ In
\cite{Bahrami} a similar derivation has been presented for the case of Newtonian gravity.}:
 \begin{equation}
\displaystyle i\frac{\partial}{\partial t}\psi(\vec{x},t)=\left[-\frac{1}{2m}\nabla^{2}\right.-Gm^{2} \displaystyle \int d^{3}x' \displaystyle \left.\frac{{\rm Erf}\left(\frac{M_{s}}{2}\left|\vec{x}'-\vec{x}\right|\right)}{\left|\vec{x}'-\vec{x}\right|}\left|\psi(\vec{x}',t)\right|^{2}\right]\psi(\vec{x},t),
\label{eq:10}
\end{equation}
where $\psi(\vec{x},t)$ is the one particle wave-function and we have assumed that only one kind of particle is present, $\hat{\rho}=m\hat{\psi}^{\dagger}\hat{\psi}$. Thus, Eq. (\ref{eq:10}) describes the dynamics of a self-gravitating wave-function in a non-relativistic, weak-field regime. Note that the non-linearity in the Schr\"odinger equation appears only at the level of the potential. We can also split the potential in two parts in such a way to isolate the main contribution due to IDG:
\begin{equation}
\Phi[\psi](\vec{x}) \simeq \displaystyle -\frac{G m^2 M_s}{\sqrt{\pi}} \int\limits_{|\vec{x}'|<2/M_s}d^3 x' \left|\psi(\vec{x}',t)\right|^2-Gm^2 \displaystyle \int\limits_{|\vec{x}'|\geq 2/M_s} d^3 x' \frac{\left|\psi(\vec{x}',t)\right|^2}{|\vec{x}'-\vec{x}|}.
\label{decomp}
\end{equation}
We can clearly notice that the first term in Eq. \eqref{decomp} is purely due to IDG, while the second one takes into account the Newtonian self-interaction as in the well-known Schr\"odinger-Newton equation (see \cite{Diosi:1988uy,Penrose,Bahrami}). Depending on the behavior of the wave-function $\psi(\vec{x},t)$, one of the two terms could give a much greater contribution compared to the other. For example, if $\sigma$ is a characteristic width such that $\psi(\vec{x},t)\rightarrow 0$ when $|\vec{x}|>\sigma$, then in the regime $\sigma < 2/M_s$, using also the fact that the wave-function is normalized to one, the second term is negligible and thus the potential tends to be constant. In such a regime, where non-locality becomes relevant, the linearity is restored.

\subsection{Mean-field approximation}

In the previous subsection the coupling to the expectation value of the energy-momentum tensor was considered, $\left\langle \psi\left|\hat{\tau}_{\mu\nu}\right|\psi\right\rangle$, and the metric perturbation $h_{\mu \nu}$ was assumed to be a classical field. We now want to study the case in which gravity is directly coupled to the quantum energy-momentum tensor so that, not only matter, but also the graviton field has to be quantized, $h_{\mu \nu} \rightarrow {\hat h}_{\mu \nu}$. In such a scenario, the starting point is now the following equations:
\begin{equation}
{\hat P}_{\mu \nu}\equiv{\hat G}_{\mu \nu}+{\hat G}^{(q)}_{\mu \nu}=8\pi G \hat{\tau}_{\mu\nu},\label{eq:3.2}
\end{equation}
where ${\hat P}_{\mu \nu}$ still arises from the equations of motion of IDG action, as shown in Ref. \cite{Biswas:2013cha}, but here it is an operator due to that fact that the metric field has to be quantized, $g_{\mu \nu}\rightarrow \hat{g}_{\mu \nu}$.  By working again in the non-relativistic regime, the analog of Eq. \eqref{eq:4} is given by
\begin{equation}
a(\boxempty_{s})\boxempty {\hat h}_{\mu\nu}={ -16\pi G\left(\hat{\tau}_{\mu\nu} -\frac{1}{2}\eta_{\mu\nu}\hat{\tau} \right)},
\label{eq:4.2}
\end{equation}
while Eq. \eqref{eq:7} becomes
\begin{equation}
a\left(\nabla_{s}^{2}\right)\nabla^{2}{\hat \Phi}=4\pi G \hat{\tau}_{00} ,\label{eq:7.2}
\end{equation}
whose solution is
\begin{equation}
{\hat \Phi}(\vec{x})=-G\int d^{3}x'\frac{{\rm Erf}\left(\frac{M_{s}}{2}\left|\vec{x}'-\vec{x}\right|\right)}{\left|\vec{x}'-\vec{x}\right|} \hat{\rho}(\vec{x}') ,\label{eq:8.2}
\end{equation}
where we have used again $\hat{\tau}_{00}=\hat{\rho}$.
One can immediately notice that it is the quantized energy-momentum tensor that now appears and not its expectation value. Moreover, the interaction Hamiltonian is now given by
\begin{equation}
\begin{array}{rl}
\hat{H}_{\text{int}}= & \displaystyle -G\int \!d^{3}xd^{3}x'\frac{{\rm Erf}\left(\frac{M_{s}}{2}\left|\vec{x}'-\vec{x}\right|\right)}{\left|\vec{x}'-\vec{x}\right|}\hat{\rho}(\vec{x}')\hat{\rho}(\vec{x})\\
= & \displaystyle -G m^2 \int d^{3}xd^{3}x'\frac{{\rm Erf}\left(\frac{M_{s}}{2}\left|\vec{x}'-\vec{x}\right|\right)}{\left|\vec{x}'-\vec{x}\right|}\hat{\psi}^{\dagger}(\vec{x}')\hat{\psi}(\vec{x}')\hat{\psi}^{\dagger}(\vec{x})\hat{\psi}(\vec{x}).
\end{array}
\label{eq:9.2}
\end{equation}
Unlike Eq. \eqref{eq:9}, the interaction Hamiltonian in Eq. \eqref{eq:9.2} is not expressed in terms of the expectation value of the density operator, and this is a crucial difference that in the case of quantized gravity will not introduce any non-linearities in the Schr\"odinger equation for an $N$-particle state. Indeed, the wave function $\Psi_{{\scriptscriptstyle N}} (\vec{x}_{{\scriptscriptstyle 1}} ,\dots,\vec{x}_{{\scriptscriptstyle N}} ,t)$ associated to an $N$-particle state satisfies the following linear Schr\"odinger equation:
\begin{equation}
i\frac{\partial}{\partial t} \Psi_N={\hat H}_{\scriptscriptstyle N}\Psi_{{\scriptscriptstyle N}}, \label{schr_N}
\end{equation}
where 
\begin{equation}
{\hat H}_{\scriptscriptstyle N}=- \sum_{i=1}^{N} \frac{\nabla^{2}_i}{2m}-\sum_{i\neq j}^{N} {\hat \Phi}_{ij} \label{H_N}
\end{equation}
is the total Hamiltonian, and
\begin{equation}
{\hat \Phi}_{ij}\equiv-Gm^2 \frac{{\rm Erf}\left(\frac{M_{s}}{2}|\hat{\vec{x}}_i-\hat{\vec{x}}_j|\right)}{|\hat{\vec{x}}_i-\hat{\vec{x}}_j|}
\end{equation}
represents the mutual gravitational potential between two components, $i$ and $j$, of the $N$-particle system; we have assumed that all $N$ masses are equal.

We are now interested in the limit $N \rightarrow \infty$ of Eq. \eqref{schr_N}, namely we want to apply a mean-field approximation. Given a one-particle state $ \left|\psi \right\rangle$, let us consider an initial factorized $N$-particle state defined as
\begin{equation}
\left|\Psi \right\rangle:=\left|\psi \right\rangle \otimes \cdots \otimes \left|\psi \right\rangle\equiv \otimes^{\scriptscriptstyle N}_{{\scriptscriptstyle i=1}}\left|\psi \right\rangle. \label{initial-factorized}
\end{equation}
Then, one can show that at later times the state in Eq. \eqref{initial-factorized} evolves as\footnote{For more details see Ref. \cite{bardos1} where the authors have applied the same procedure to the case of $N$ bosons field interacting through Coulomb potential in a mean-field regime, and Ref. \cite{hu} for a similar derivation in the case of Newtonian gravitational interaction. The rigorous mathematical derivation requires the so called quantum Bogoliubov-Born-Green-Kirwood-Y von hierarchy in the large $N$-limit \cite{bardos}.}
\begin{equation}
\lim\limits_{{\scriptscriptstyle N\rightarrow \infty}} e^{i{\hat H}_{\scriptscriptstyle N}t} \otimes^{\scriptscriptstyle N}_{{\scriptscriptstyle i=1}}\left|\psi \right\rangle=\left|\psi(t) \right\rangle,
\end{equation}
where ${\hat H}_{\scriptscriptstyle N}$ is the Hamiltonian defined in Eq. \eqref{H_N} and the wave-function $\psi(\vec{x},t)$, corresponding to the state $\left|\psi(t) \right\rangle$, satisfies the same integro-differential equation that we have derived in Eq. \eqref{eq:10}. Note that in this case $\psi(\vec{x},t)$ is $\it{not}$ a one-particle wave-function, but describes a many-particle system, with a large number of components $(N\rightarrow \infty)$, i.e. a condensate.

We can now summarize the results obtained in this section: in both cases of semi-classical approach and quantum gravitational interaction the dynamics of a self-gravitating system in the non-relativistic regime can be mathematically described in terms of the same equation \eqref{eq:10}. However, one has to keep in mind that in the case of classical gravity Eq. \eqref{eq:10} can describe the dynamics of one single particle or molecule, while in the case of quantized gravity the same equation is seen as a Hartree equation in the mean-field approximation, and takes into account a collective dynamics, where the self-potential term is due to the sum of all mutual gravitational interaction-contributions among all particles of the system.  

\section{Quantum solitonic solutions}

Let us now study the wave-function of the ground-state of the system, namely seeking solutions which balance quantum-mechanical spreading and contraction due to the attractiveness of the gravitational interaction. Since the mathematical form of Eq. \eqref{eq:10} is valid either for classical and quantized gravity, we can perform a unique analysis that includes both cases. The two different interpretations will have to be distinguished in the moment in which one wants to question the experimental testability of the models, as we will see in the next section.

\begin{figure}[t]
	\centering
	\includegraphics[scale=0.45]{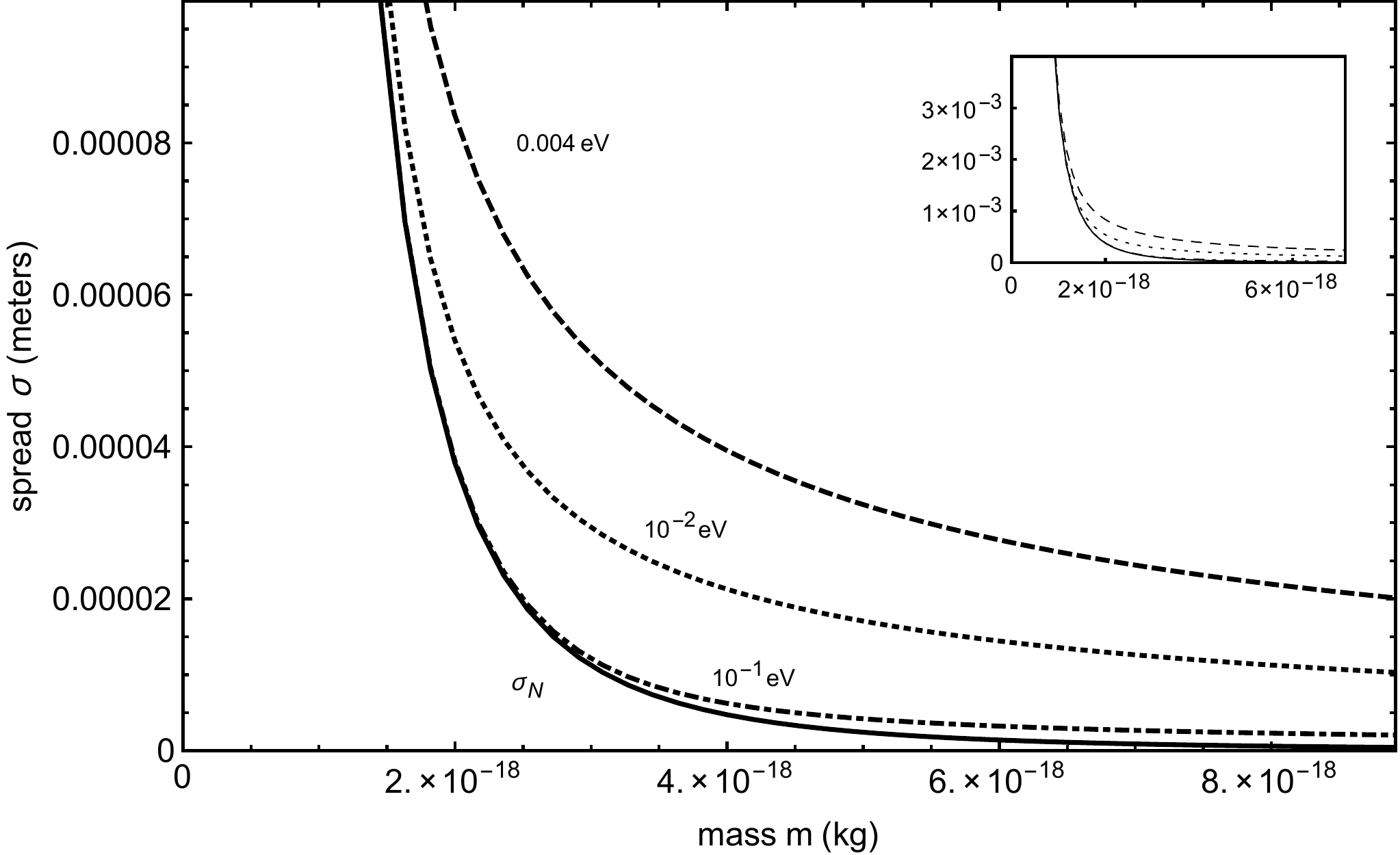}
	\protect\caption{In the above plot we have shown the quantum spreads for  the Newtonian potential,
		$\sigma_{{\scriptscriptstyle N}}$, by the solid line, and for IDG potential, $\sigma_{{\scriptscriptstyle IDG}}$,
		with respect to the mass, $m$ (in kg). For the IDG case, we have also considered different values of $M_s= 0.004\,\text{eV}$ (dashed line), 
		$10^{-2}\,\text{eV}$ (dotted line) and $10^{-1}\,\text{eV}$ (dot-dashed line). The lowest value of $M_s$ arises from the table-top experiment,
		which has seen no departure from Newtonian $1/r$-potential~\cite{-D.-J.}. It is evident from the smaller plot, in which a wider range of values of the spread has been considered, that once $M_{s}$ is
		fixed, $\sigma_{{\scriptscriptstyle IDG}} \longrightarrow \sigma_{{\scriptscriptstyle N}}$ for $\sigma >2/M_s$. }	\label{fig1}
\end{figure}

Following Ref. \cite{Diosi:1988uy}, we can find approximate soliton-like solutions of minimal energy for the ground-state of the form $\varphi(\vec{x},t)=\varphi(\vec{x})e^{-i\epsilon t}$, where  $\varphi(\vec{x})$ is a real function depending only on the space-coordinates $\vec{x}$, while the time-dependence only appears in the phase, and $\epsilon$ is the Lagrange multiplier arising from minimising the energy of the ground-state. By assuming that $\varphi(\vec{x})$ is a Gaussian wave-packet, with a normalized 
$\varphi(\vec{x})$, and  with a characteristic width $\sigma$, i.e.
\begin{equation}
\varphi(\vec{x})={ \frac{1}{\pi^{3/4}\sigma^{3/2}}e^{-\left|\vec{x}\right|^{2}/2\sigma^{2}}},\,\,\,\,\,\,\,\,\,\,\int d^{3}x\varphi(\vec{x})^{2}=1,\label{eq:11}
\end{equation}
we obtain the expectation value of the energy in the solitonic ground-state:
\begin{equation}
\begin{array}{rl}
E_{{\scriptscriptstyle IDG}}= & \displaystyle \int d^{3}x\varphi(\vec{x})\left[-\frac{1}{2m}\nabla^{2}- Gm^{2}  \int d^{3}x'\frac{\varphi^{2}(\vec{x}')}{\left|\vec{x}'-\vec{x}\right|} {\rm Erf}\left(\frac{M_{s}}{2}\left|\vec{x}'-\vec{x}\right|\right)\right]\varphi(\vec{x})\\
= &  \displaystyle  \frac{3}{4}\frac{1}{m\sigma^{2}}-\sqrt{\frac{2}{\pi}}Gm^{2}\frac{M_{s}}{\sqrt{2+M_{s}^{2}\sigma^{2}}}.
\end{array}
\label{eq:12}
\end{equation}
Note that in the limit $M_{s}\sigma>2$, we recover the energy in the case of Newtonian self-interaction (see Ref. \cite{Diosi:1988uy}), $E_{{\scriptscriptstyle N}}=\frac{3}{4}\frac{1}{m\sigma^{2}}-\sqrt{\frac{2}{\pi}}\frac{Gm^{2}}{\sigma},$ as expected. 
Further note,  that the presence of $M_s$ increases the energy of the ground state, in fact for any values of $\sigma,$ $m$ and $M_s$, we find:
\begin{equation}
 E_{{\scriptscriptstyle IDG}}\geq E_{{\scriptscriptstyle N}}\,.
\end{equation}
Now, we can also find the spread $\sigma_{{\scriptscriptstyle IDG}}$ of the soliton by minimising the energy in Eq. (\ref{eq:12}) with respect to 
$\sigma$ and we obtain:
\begin{equation}
-\frac{3}{2m\sigma_{{\scriptscriptstyle IDG}}^{3}}+\sqrt{\frac{2}{\pi}}Gm^{2}M_{s}^{3}\frac{\sigma_{{\scriptscriptstyle IDG}}}{\left(2+M_{s}^{2}\sigma_{{\scriptscriptstyle IDG}}^{2}\right)^{3/2}}=0.\label{eq:11.eq}
\end{equation}
It is not instructive to write down the full physical solution $\sigma_{{\scriptscriptstyle IDG}}$ here, we will provide its full behaviour in Fig.~\ref{fig1}.
Note that in the limit $M_{s}\sigma>2$, we recover the Newtonian case already, seen in Eq.~(\ref{eq:13}), obtained in Ref. \cite{Diosi:1988uy}.
In the opposite limit, when $M_s\sigma <2$, by expanding the energy in Eq. \eqref{eq:12}, up to the order $\mathcal{O}(M_s^{3})$, and simplifying $\frac{M_s}{\sqrt{2+M_{s}^{2}\sigma^{2}}}\simeq\frac{M_s}{\sqrt{2}}-\frac{M_{s}^{3}\sigma^{2}}{4\sqrt{2}},$ we find
\begin{equation}
\begin{array}{rl}
\sigma_{{\scriptscriptstyle IDG}}^{{\scriptscriptstyle (M_{s}\sigma<2)}}\simeq & \displaystyle \left(3\sqrt{\pi}\right)^{\frac{1}{4}}\left(\frac{1}{Gm^{3}}\right)^{\frac{1}{4}}\left(\frac{1}{M_{s}}\right)^{\frac{3}{4}}\\
= & \displaystyle \left(2\sqrt{2}\right)^{\frac{1}{4}}\sigma_{{\scriptscriptstyle N}}^{\frac{1}{4}}\left(\frac{1}{M_{s}}\right)^{\frac{3}{4}}.
\end{array}\label{eq:14}
\end{equation}
The above equation suggests that the quantum state of a self-gravitating system is localized within a region of size $1/M_{s}$, which is governed by the scale of non-locality, this will become evident by looking at the plot below in Fig.~\ref{fig1}, and at the table $\mathrm{I}$ in which several values of the spread of the wave-function have been shown.

In Fig.~\ref{fig1}, the physical solution of Eq. \eqref{eq:11.eq} has been plotted, i.e. the spread of the solitonic wave-packet with respect to the mass for different values of the parameter $M_{s}$. The quantum spread $\sigma_{{\scriptscriptstyle IDG}}$ and $\sigma_{{\scriptscriptstyle N}}$ are plotted as functions of the mass $m$. Note  that in the case of IDG, the wave-function spread turns out to be much larger than that of the Newtonian gravitational potential, 
\begin{equation}
\sigma_{{\scriptscriptstyle IDG}}\geq \sigma_{{\scriptscriptstyle N}}.
\end{equation}

\begin{table*}[b]
	\caption{Values of the spreads $\sigma_{{\scriptscriptstyle N}}$ and $\sigma_{{\scriptscriptstyle IDG}}$ of the quantum wave-function (in meters) for fixed values of the mass $m$ (in kg) and for sample values of
		$M_s=0.004{\rm eV}, 1{\rm eV}, 1{\rm GeV}, 10^{5}{\rm GeV}$.}
	\centering
	\begin{tabular}{p{0.10\linewidth}p{0.125\linewidth}p{0.155\linewidth}p{0.155\linewidth}p{0.155\linewidth}p{0.16\linewidth}}
		\toprule[1pt]\midrule[0.3pt]
		mass (kg)  & $\sigma_{{\scriptscriptstyle N}}$(m) & $\sigma_{{\scriptscriptstyle IDG}}(0.004$eV)(m)  & $\sigma_{{\scriptscriptstyle IDG}}(1$eV)(m) & $\sigma_{{\scriptscriptstyle IDG}}(1$GeV)(m) & $\sigma_{{\scriptscriptstyle IDG}}(10^{5}$GeV)(m)\\
		\midrule[0.3pt]
		$10^{-10}$ & $3.02\times10^{-28}$ & $1.00\times10^{-10}$ & $1.60\times10^{-12}$ & $2.85\times10^{-19}$  &  $5.06\times10^{-23}$\\
		$10^{-12}$ & $3.02\times10^{-22}$ & $3.18\times10^{-9}$ & $5.06\times10^{-11}$ & $9.00\times10^{-18}$ &  $1.83\times10^{-21}$\\
		$10^{-14}$ & $3.02\times10^{-16}$ & $1.01\times10^{-7}$ & $1.60\times10^{-9}$ & $4.73\times10^{-16}$ &  $3.02\times10^{-16}$ \\
		$10^{-16}$ & $3.02\times10^{-10}$ & $3.18\times10^{-6}$ & $5.12\times10^{-8}$ & $3.02\times10^{-10}$  & $3.02\times10^{-10}$\\
		\midrule[0.3pt]\bottomrule[1pt]
	\end{tabular}
\end{table*}
This is an effect induced by IDG, and the presence of non-locality in the gravitational interaction. Such theories possess mass-gap~\cite{Frolov}
in the gravitational potential, governed by the scale $M_s$, and therefore within $\sigma< 2/M_s$, effectively the gravitational force vanishes, but for  $\sigma > 2/M_s$, the gravitational attractive force balances the quantum spread of the wave-function. In this regard, the quantum wave-function of the ground-state of the meso-scopic system can be described in terms of a solitonic solution, as depicted in Fig.~\ref{fig1}. In the plot-region $\sigma < 2/M_s$ the behaviour of the IDG spread is well approximated by Eq. \eqref{eq:14}. Moreover, the larger is the value of $M_{s}$, smaller is the value of 
$\sigma_{{\scriptscriptstyle IDG}}$, and in the limit in which $\sigma> 2/M_s$, we have 
$\sigma_{{\scriptscriptstyle IDG}}\rightarrow\sigma_{{\scriptscriptstyle N}}$. Generally, we obtain: $\sigma_{{\scriptscriptstyle IDG}}^{{\scriptscriptstyle (M_{s}=0.004\,\text{eV})}}\leq\sigma_{{\scriptscriptstyle IDG}}\leq\sigma_{{\scriptscriptstyle IDG}}^{{\scriptscriptstyle (M_{s}=\infty)}}=\sigma_{{\scriptscriptstyle N}}$.

\section{Discussion and conclusions}
In the previous section, by minimizing the energy in Eq. \eqref{eq:12}, we have found approximate solitonic solutions for the ground-state of a self-gravitating system in the non-relativistic regime, where the gravitational interaction is described by IDG. 
In the table $\mathrm{I}$, we consider some numerical values of the spread of the wave-function obtained by evaluating the physical solution of Eq. \eqref{eq:11.eq} for different values of the mass, $m$, and the parameter $M_{s}$. Note that the larger the mass is, smaller is the spread. Clearly, as we have already emphasized above, in the semi-classical approach $m$ represents the mass of one single atom or molecule, while in the case of quantized gravity it corresponds to the total mass of a condensate. For a mass $m=10^{-14}$ kg, in the Newtonian case, 
$\sigma_{{\scriptscriptstyle N}}=3.02\times10^{-16}$ meters, while the IDG spread is always larger, and for $M_s=0.004$eV,   
$\sigma_{{\scriptscriptstyle IDG}}=1.01\times10^{-7}$ meters, which is much larger than $\sigma_{{\scriptscriptstyle N}}$. Such a gravitational-induced effect on the dynamics of the wave-packet, not only would offer a new framework in which one can investigate more deeply the real nature of gravity, i.e. whether it is classical or quantum, but would also open a new window of opportunity to test short-distance gravity beyond Einstein's GR.

In order to discuss the experimental testability of the these models, we first need to specify how to treat gravity, either classical or quantum. In fact, there are experiments where the low-energy system is composed by a very small number of atoms, and others in which one deals with many-particle systems with a large number of components, for instance condensates. The former case would apply to the case of classical gravity coupled to quantum matter through a semi-classical approach, while the latter would be suitable to test the quantum properties of the gravitational interaction.

This work {\it indeed} provides a new scenario for testing the classical properties of IDG from the quantum localisation of the wave-packet in molecule-interferometry \cite{gerlich}\footnote{See also Refs. \cite{yang,grobardt-bateman,andregro} for possible optomechanical tests aimed to test semi-classical gravity.}. A larger spread in the wave-function might provide us a smoking gun signature of the nature of the gravitational potential. We should be able to study the free expansion and contraction of the wave-function and place limits on the new scale of physics, $M_s$. It is worthwhile to mention that these are very sensitive experiments, and there are several sources of decoherence effects, for a review see~\cite{Bassi}, which pose enormous experimental challenges for observing some interesting quantum phenomena in a ghost free and singularity free theory of gravitation.

Moreover, one promising experiment in which quantum properties of gravity might be tested could be tests performed in microgravity, see~\cite{Muntinga:2013pta}, where the authors have put forward an interesting experimental proposal to test quantum mechanics of weakly coupled Bose-Einstein condensate (BEC) in a freely falling system. In this case one is interested in studying interference for a free-falling BEC, that is treated as a many-particle system and thus Eq. \eqref{eq:10} has to be interpreted as a Hartree equation in the mean-field appproximation. Such an experiment is progressing and in future we might be able to constrain the scale of non-locality $M_s$ by comparing our predictions with the experimental data. It is worth mentioning that the BEC considered in Ref. \cite{Muntinga:2013pta} is made of about  $10^4$ atoms, and in the near future one could be able to create quantum interference with $10^6$, or even more, atoms. 

Before we conclude, it is worthwhile to mention that it is also possible to study  modified gravitational potentials, such as Yukawa-like potential \cite{capozziello} of the form $\Phi(r)=-Gm/r(1+1/3\text{exp}(-\mu r))$, arising in $f({\mathcal R})$ theories, where similar computations suggest an opposite scenario compared to that of IDG. Indeed, the energy and the spread of the ground-state turn out to be smaller compared to that of the Newtonian case: $\sigma_{{\scriptscriptstyle f(\mathcal{R})}}\leq \sigma_{{\scriptscriptstyle N}}\leq \sigma_{{\scriptscriptstyle IDG}}$.

This proof-of-concept paper provides us, for the first time, how we can put theories of gravity on to test-bed by studying the solitonic wave-function of a self-gravitating quantum system
in theories beyond Einstein's GR. In particular, singularity free theories of gravity provides an intriguing observation that the minimum energy for  the
IDG is always larger compared to that of the Newtonian case, and the spread of the wave-function is always larger than that of the Newtonian potential for meso-scopic systems. 
These predictions are indeed testable in a table-top experiment in the near future, furthermore we can possibly constrain, $M_s$,  allowing us a deeper understanding of classical and quantum properties of the gravitational interaction at short distances.

\section*{Acknowledgments}
The authors would like to thank Sougato Bose for discussions.

\end{document}